# Introducing MISS, a new tool for collision avoidance analysis and design

Juan Luis Gonzalo[a,*], Camilla Colombo[a], and Pierluigi Di Lizia[a]

[a] Politecnico di Milano, Via la Masa 34, 20156 Milan, Italy.
[*] Corresponding author, juanluis.gonzalo@polimi.it

## ABSTRACT

The core aspects and latest developments of Manoeuvre Intelligence for Space Safety (MISS), a new software tool for collision avoidance analysis and design, are presented. The tool leverages analytical and semi-analytical methods for the efficient modelling of the orbit modifications due to different control strategies, such as impulsive or low-thrust manoeuvres, and maps them into displacements at the nominal close approach using relative motion equations. B-plane representations are then used to separate the phasing-related and geometry-related components of the displacement. Both maximum miss distance and minimum collision probability collision avoidance manoeuvres are considered. The tool also allows for the computation of state transition matrices and propagation of uncertainties. Several test cases are provided to assess the capabilities and performance of the tool.

## 1 INTRODUCTION

Technical and organizational requirements for Space Situational Awareness (SSA) and collision avoidance activities are steadily increasing alongside the growing use of space-based assets. This affects a wide range of topics, from policy aspects such as the new guidelines under development by the Inter-Agency Space Debris Coordination Committee, to the continuous improvement and introduction of SSA capabilities like the Space Fence. Recently declared operational, this new tracking and monitoring system developed by Lockheed Martin for the US Air Force will improve accuracy and allow to follow objects significantly smaller than the previous 10 cm limit. From the operational point of view, recent reductions in the cost of access to space, spurred by increased competition in the launch market and new cost-effective platforms, together with the proposed deployment of large constellations, will naturally lead to an increase in the number of close approaches (CA). In this context, operators will not only have to decide if and how to perform a collision avoidance manoeuvre (CAM), but also do so minimizing the risk of additional CAs down the road. An additional degree of flexibility (and complexity) comes from the growing variety of propulsion systems. Many current satellites complement or substitute traditional impulsive thrusters with low-thrust electric propulsion systems or others like sails and tethers. Put together, these elements define an increasingly complex scenario for satellite operators, who require fast and reliable tools for the analysis of potential conjunctions and the design of CAMs.

Gathering recent advances in CAM analysis and design [1][2][3], the Manoeuvre Intelligence for Space Safety (MISS) software tool has been introduced to provide an integrated approach to SSA-related operational activities. MISS is being developed within the European Research Council-funded project COMPASS [4], which studies orbital perturbations and how they can be leveraged for mission design. The tool provides a holistic approach to CAM design, considering both impulsive and continuous, low-thrust manoeuvres (including not only electric thrusters, but also other devices such as sails). Several design strategies are considered, mainly maximizing miss distance or minimizing collision risk, and physically significant interpretations of the CAM are provided by leveraging the b-plane representation. Analytical and semi-analytical (SA) techniques are exploited to provide a fast and flexible tool. Particularly, the use of State Transition Matrices (STMs) allows for the fast propagation of uncertainties and to perform extensive sensitivity analyses over different parameters of the CA and the spacecraft (e.g. area-to-mass ratio or drag and reflectivity coefficients). Whenever possible, STMs are computed analytically; in the rest of cases, they are obtained through the semi-analytical propagation of the single-averaged variational equations using PlanODyn [5].

The manuscript is organized as follows. First, the three main components of MISS are introduced, and the underlying physical models are presented. The first component deals with the modelling of orbital elements modification using analytical and SA approaches, the second one covers the use of relative motion equations to map the changes in orbital elements into displacements at the CA, and the last one allows for the analysis of the results through their projection onto the b-plane and computation of the collision probability. The next section presents some practical application cases. Finally, the capabilities and performance of the tool are assessed through several test cases, and conclusions are drawn.

## 2 DYNAMICAL MODELS AND SOFTWARE TOOL STRUCTURE

Let us consider a CA at a time $t_{CA}$ between a manoeuvrable spacecraft and a debris. The debris denomination is used here in a broad sense, referring to an object not collaborating in the CAM. For simplicity, in the following the nominal miss distance at $t_{CA}$ is supposed to be zero (direct impact). The core aim of MISS is to characterize the updated CA, both in terms of miss distance and collision probability, due to a CAM performed by the spacecraft. To do this, the software implements a computation kernel composed of three main elements, as depicted in Fig. 1:
1. Dynamical models for the change in orbital elements, based on analytical and SA methods.
2. Linearized relative motion equations around the nominal orbit of the manoeuvring spacecraft, relating changes in the orbital elements to displacements (and relative velocities) at the CA.
3. B-plane projection and collision probability computation.

The following subsections introduce each one of these elements and their underlying physical models.

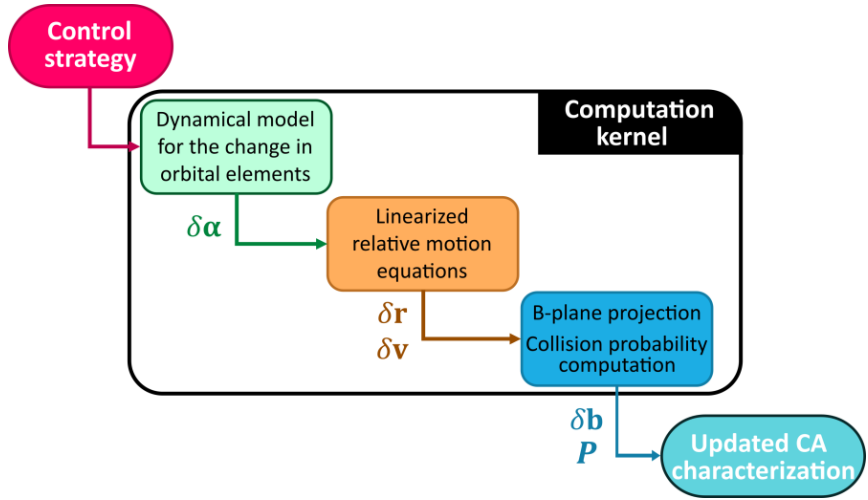

Fig. 1 Schematic representation of the main elements in the computation kernel of MISS.

### 2.1 Change in orbital elements

The change in orbital elements is modelled in all cases through Gauss' planetary equations [6]:

$$\begin{aligned}
\frac{da}{dt} &= \frac{2a^2 v}{\mu} a_t, & \frac{de}{dt} &= \frac{1}{v}\left[2(e + \cos f) a_t - \frac{r}{a}\sin f\, a_n\right] \\
\frac{di}{dt} &= \frac{r \cos \theta}{h} a_h, & \frac{d\Omega}{dt} &= \frac{r \sin \theta}{r \sin i} a_h \\
\frac{d\omega}{dt} &= \frac{1}{ev}\left[2 \sin f\, a_t + \left(2e + \frac{r}{a}\cos f\right) a_n\right] - \frac{r \sin \theta \cos i}{h \sin i} a_h \\
\frac{dM}{dt} &= n - \frac{b}{eav}\left[2\left(1 + \frac{e^2 r}{p}\right)\sin f\, a_t + \frac{r}{a}\cos f\, a_n\right]
\end{aligned} \qquad (1)$$

where $\boldsymbol{\alpha} = [a, e, i, \Omega, \omega, M]$ are the classic orbital elements (semi-major axis, eccentricity, inclination, right ascension of the ascending node, argument of pericentre, and mean anomaly, respectively), $f$ is the true anomaly, $\theta$ is the argument of latitude, and $a_t$, $a_n$, $a_h$ are the tangential, normal, and out-of-plane components of the perturbing acceleration.

For an impulsive CAM at a time $t_{CAM}$, the change in orbital elements can be written linearly as [7][1][2]:

$$\delta\boldsymbol{\alpha}(t_{\text{CAM}}) = \mathbf{G}_v(t_{\text{CAM}})\delta\boldsymbol{v}(t_{\text{CAM}}) \qquad (2)$$

where the elements of $\mathbf{G}_v(t_{\text{CAM}})$ can be obtained directly from Eq. (1) as the coefficients of the corresponding components of the acceleration. Furthermore, it is possible to extend the model to account also for a displacement $\delta\boldsymbol{r}(t_{\text{CAM}})$ [1][2]:

$$\delta\boldsymbol{\alpha}(t_{\text{CAM}}) = \begin{bmatrix} \mathbf{G}_r(t_{\text{CAM}}) \\ \mathbf{G}_v(t_{\text{CAM}}) \end{bmatrix} \delta\mathbf{s}(t_{\text{CAM}}) = \mathbf{G}(t_{\text{CAM}})\delta\mathbf{s}(t_{\text{CAM}}) \qquad (3)$$

where $\delta\mathbf{s} = [\delta\mathbf{r}\ \delta\mathbf{v}]$ is the change in the Cartesian state at $t_{\text{CAM}}$. Although this extended expression is not needed for the computation of CAMs, it will be useful for the construction of STMs. The derivation and final expressions of $\mathbf{G}_r(t_{CAM})$ are rather lengthy and are omitted for conciseness; they can be found in [1][2]. Suffice to say that it has a structure analogous to $\mathbf{G}_v(t_{\text{CAM}})$.

The change in $M$ given by Eq. (2) is not enough to fully characterized the new mean anomaly at $t_{\text{CAM}}$. Certainly, as detailed in [7], the change in mean motion due to the CAM also has to be accounted. Then, it is possible to write:

$$\delta M(t_{\text{CA}}) = \delta M(t_{\text{CAM}}) + \delta n\, \Delta t = \delta M(t_{\text{CAM}}) - \frac{3}{2}\frac{\sqrt{\mu}}{a^{5/2}}\Delta t\, \delta a \qquad (4)$$

where $\Delta t = t_{\text{CAM}} - t_{\text{CA}}$ is the lead time of the CAM. Note that the convention used in this paper differs slightly form the one in [7][1], where the additional term for the mean anomaly was treated as a contribution by $\delta a$ and integrated in the matrix for the linearized relative motion. Although that approach has the advantage of providing a simpler expression for $\delta\boldsymbol{\alpha}$, considering only its value at $t_{\text{CAM}}$ instead of both at $t_{\text{CAM}}$ and $t_{\text{CA}}$, it has the drawback of relying on a modified equation for the relative motion. In order to properly separate the computation of the change in orbital elements from the application of the relative motion equations and to improve compatibility with other application cases (such as the low-thrust model), the change of orbital elements at $t_{\text{CA}}$ is also introduced in this work, noting that $\delta\boldsymbol{\alpha}(t_{\text{CA}}) = \delta\boldsymbol{\alpha}(t_{\text{CAM}})$ except for $\delta M(t_{\text{CA}})$, which has to be computed according to Eq. (4).

A SA model for a low-thrust CAM is also considered. In this case, the control law consists of a constant, tangential thrust applied between a time $t_i$ and a time $t_e$ before the CA. The choice of tangential thrust is based on the results for the impulsive CAM, which show that the optimal CAM tends to align with the tangential direction for lead times greater than a period of the manoeuvring spacecraft, as seen in [7][1] and the test cases in Section 4. A set of SA expressions for the evolution of the mean values of all orbital elements except $M$ is obtained by introducing $a_n = a_h = 0$ in Eq. (1), expressing them in terms of the eccentric anomaly using the approximate relation [8][9][3]:

$$\frac{dE}{dt} \approx \sqrt{\frac{\mu}{a^3}}\frac{1}{1 - e\cos E} \qquad (5)$$

and integrating over a whole revolution of $E$ (assuming that all elements remain constant). The resulting expressions for the mean variations of $a$ and $e$ during a complete revolution are:

$$\Delta a = a_t \frac{2a^3\eta}{\mu}\, \mathrm{E}\left[E, -\frac{e^2}{1-e^2}\right]_{E=0}^{E=2\pi}$$

$$\Delta e = a_t \frac{2a^2\eta^2}{\mu e}\left[\frac{1}{2}\ln\left(\frac{\sqrt{1-e^2\cos^2 E}+e\sin E}{\sqrt{1-e^2\cos^2 E}-e\sin E}\right) - \frac{1}{\eta}\mathrm{F}\left[E, -\frac{e^2}{1-e^2}\right] + \eta\, \mathrm{E}\left[E, -\frac{e^2}{1-e^2}\right]\right]_{E=0}^{E=2\pi} \qquad (6)$$

where $\eta = \sqrt{1-e^2}$, $\mathrm{E}[\cdot,\cdot]$ is the incomplete elliptic integral of the first kind, and $\mathrm{F}[\cdot,\cdot]$ is the incomplete elliptic integral of the second kind. On the other hand, $\Delta\omega$ turns out to be zero, and $i$ and $\Omega$ will remain constant for all $E$ as there is no thrust component in the out-of-plane direction. The total changes $\delta a$ and $\delta e$ after a complete number of revolutions can be computed by sequentially evaluating $\Delta a$ and $\Delta e$ for each revolution (updating the value of $e$), or approximated simply by multiplying $\Delta a$ and $\Delta e$ by the number of revolutions. In case the time span $[t_i, t_e]$ corresponds to an incomplete number of revolutions, the short-periodic terms are approximated as:

$$x(E) = x_0 + \frac{\Delta x}{2\pi}(E - E_0) + K_x(\cos(E - \varphi_x) - \cos(E_0 - \varphi_x)) \qquad (7)$$

where $x$ represents $a$, $e$, or $\omega$ (note that, although the mean variation of $\omega$ was zero, it does present a short-periodic evolution). Coefficients $K_x$ and $\varphi_x$ are determined numerically, performing a least-squares fit over a set of numerical values obtained by propagating one revolution of the orbit. Note that only one revolution has to be propagated, and the resulting coefficients are valid regardless of $t_i$ and $t_e$ (but depend on $a_t$).

Same as with the impulsive case, the determination of $\delta M$ entails additional difficulties. Moreover, because $\delta M$ determines the change of phasing at the CA, it strongly affects the accuracy of the results. Colombo et al. [10] show that $\delta M$ can be expressed as:

$$\delta M = (n_e - n_i)t_{CA} + n_i t_i - n_e t_e + \Delta M \tag{8}$$

where $n_i$ and $n_e$ are the mean motions at $t_i$ and $t_e$, respectively. Analogously to the impulsive case, the first part represents the change in $M$ due to the change in mean motion, while $\Delta M$ corresponds to the change in $M$ directly introduced by the CAM. Note that computing $\Delta M$ for a whole number of revolutions in the eccentric anomaly will yield $\Delta M = 0$, as they both have the same orbital period. However, $\Delta M$ can vary significantly for incomplete revolutions due to the contribution of the periodic terms. In order to retrieve this effect, the same scheme proposed in [10] is applied. First, the exact final eccentric anomaly is obtained by integrating the time law, Eq.(5), numerically between $t_i$ and $t_e$, using the short-periodic expressions given by Eq.(7). The value of $\Delta M$ is then computed integrating the differential equation for the mean anomaly:

$$\frac{dM}{dE} = (1 - e\cos E)\left(1 - a_t \frac{2a^2(1 - e^3 \cos E)\sin E}{e\mu\sqrt{1 - e^2 \cos^2 E}}\right) \tag{9}$$

only over the last, incomplete revolution. Again, the instantaneous evolutions of $a$ and $e$ are approximated through the expressions including the short-periodic corrections. The need to perform the numerical integration of the time law during the whole duration of the CAM and the numerical integration of the Gauss' planetary equation for $M$ during the last, incomplete revolution, together with the fitting process for the coefficients of the short-periodic corrections, is what prevents the method from being fully analytical. A substantial computational cost advantage is still obtained over the full integration of Gauss' equations for all elements, as will be shown in Section 4.

## 2.2 Linearized relative motion equations

The change in orbital elements is mapped to a displacement at $t_{CA}$ using a linearized relative motion model [11]:

$$\delta r_r \approx \frac{r}{a}\delta a + \frac{a\, e\, \sin f}{\sqrt{1-e^2}}\delta M - a\cos f\, \delta e$$

$$\delta r_\theta \approx \frac{r}{(1-e^2)^{3/2}}(1 + e\cos f)^2 \delta M + r\delta\omega + \frac{r\sin f}{1-e^2}(2 + e\cos f)\delta e + r\cos i\, \delta\Omega \tag{10}$$

$$\delta r_h \approx r(\sin\theta\, \delta i - \cos\theta\, \sin i\, \delta\Omega)$$

where $\delta r_r$, $\delta r_\theta$, and $\delta r_h$ are the displacements along the radial, transversal, and out-of-plane directions, respectively. Note that both $\delta \mathbf{r}$ and $\delta \boldsymbol{\alpha}$ are given at $t_{CA}$, while the values for $\boldsymbol{\alpha}$ and related magnitudes are the nominal ones (i.e. without the CAM). The model can also be extended to include the changes in relative velocity [1]:

$$\delta \mathbf{s}(t_{CA}) \approx \mathbf{A}\, \delta\boldsymbol{\alpha}(t_{CA}) = [\mathbf{A}_r\ \mathbf{A}_v]\, \delta\boldsymbol{\alpha}(t_{CA}) \tag{11}$$

where $\delta \mathbf{s} = [\delta \mathbf{r}\ \delta \mathbf{v}]$ is the change of the Cartesian state expressed in its radial, transversal, and out-of-plane components, and matrix $\mathbf{A}_v$ is reported in [1]. Note that the computation of $\delta \mathbf{v}$ is not required to analyse the CAM, but it can be used for the computation of STMs.

## 2.3 B-plane projection and collision probability computation

A convenient representation of the spacecraft's deviation can be obtained by using the b-plane, defined as the plane orthogonal to the relative velocity of the spacecraft with respect to the debris. The reference frame $\mathcal{B} = \{D; \xi, \eta, \zeta\}$ associated to the b-plane is centred at the debris $D$ and has unit vectors:

$$\mathbf{u}_\eta = \frac{\mathbf{v}_{SC} - \mathbf{v}_D}{||\mathbf{v}_{SC} - \mathbf{v}_D||}, \quad \mathbf{u}_\xi = \frac{\mathbf{v}_D \times \mathbf{u}_\eta}{||\mathbf{v}_D \times \mathbf{u}_\eta||}, \quad \mathbf{u}_\zeta = \mathbf{u}_\xi \times \mathbf{u}_\eta, \tag{12}$$

where $\mathbf{v}_{SC}$ and $\mathbf{v}_D$ are the velocities of the spacecraft and the debris, respectively. One key feature of the b-plane is that it naturally decomposes the displacements due to orbit geometry modification and due to change of phasing into the geometry axis $\xi$ and the time axis $\zeta$, respectively. The b-plane projection of $\delta\mathbf{r}$ can be expressed as [12]:

$$\delta\mathbf{b} = \mathbf{M}_{\delta b}\delta\mathbf{r}, \quad \text{with } \mathbf{M}_{\delta b} = \begin{bmatrix} \mathbf{u}_{\eta_2}^2 + \mathbf{u}_{\eta_3}^2 & -\mathbf{u}_{\eta_1}\mathbf{u}_{\eta_2} & -\mathbf{u}_{\eta_1}\mathbf{u}_{\eta_3} \\ -\mathbf{u}_{\eta_1}\mathbf{u}_{\eta_2} & \mathbf{u}_{\eta_1}^2 + \mathbf{u}_{\eta_3}^2 & -\mathbf{u}_{\eta_2}\mathbf{u}_{\eta_3} \\ -\mathbf{u}_{\eta_1}\mathbf{u}_{\eta_3} & -\mathbf{u}_{\eta_2}\mathbf{u}_{\eta_3} & \mathbf{u}_{\eta_1}^2 + \mathbf{u}_{\eta_2}^2 \end{bmatrix} \quad (13)$$

where the components of $\mathbf{u}_\eta$ must be expressed in the same reference frame as $\delta\mathbf{r}$.

Collision probability is computing following Chan's method [13]. The original conjunction in the b-plane, where both debris and spacecraft have a covariance matrix and spherical envelope, is reduced to an equivalent problem by assigning a combined covariance matrix to the debris (assumed to have no size) and a combined spherical envelope to the spacecraft (assumed to have no uncertainties). If the covariance determination for both objects is statistically independent the combined covariance can be obtained just by adding the individual covariances, while the combined spherical envelope is centred at the spacecraft and has a radius $r_A$ equal to the sum of the radii of the individual envelopes. The collision probability can then be approach by the convergent series:

$$P(u,v) = e^{-v/2} \sum_{m=0}^{\infty} \frac{v^m}{2^m m!}\left(1 - e^{-u/2}\sum_{k=0}^{m}\frac{u^k}{2^k k!}\right) \quad (14)$$

with

$$u = r_A^2/\sigma_\xi\sigma_\zeta\sqrt{1-\rho_{\xi\zeta}^2}, \quad v = \left[\left(\frac{\xi}{\sigma_\xi}\right)^2 + \left(\frac{\zeta}{\sigma_\zeta}\right)^2 - 2\rho_{\xi\zeta}\frac{\xi}{\sigma_\xi}\frac{\zeta}{\sigma_\zeta}\right]/(1-\rho_{\xi\zeta}^2) \quad (15)$$

where $(\xi,\zeta)$ is the position of the spacecraft in the b-plane. Although the number of terms for a given accuracy is not known a priory, Chan shows that accurate results can be obtained for small $u$ with as little as three terms.

## 3 APPLICATION SCENARIOS

The analytical and SA models presented in the previous section are leveraged in MISS to tackle different problems related to CAM design and CA analysis.

### 3.1 Impulsive CAM optimization

Combining Eqs. (2),(4),(11),(13), the deviation due to a CAM can be written as a linear expression on the $\delta\mathbf{v}$ imparted [1]:

$$\delta\mathbf{r} = \mathbf{T}\,\delta\mathbf{v}$$
$$\delta\mathbf{b} = \mathbf{M}_{\delta b}\mathbf{T}\delta\mathbf{v} = \mathbf{Z}\delta\mathbf{v} \quad (16)$$

Conway proved in [14] that $\|\delta\mathbf{r}\|$ is maximized by choosing a $\delta\mathbf{v}_{opt}$ parallel to the eigenvector conjugated to the maximum eigenvalue of the quadratic form $\mathbf{T}^T\mathbf{T}$. Clearly, the same result applies to $\|\delta\mathbf{b}\|$ considering the quadratic form $\mathbf{Z}^T\mathbf{Z}$. Note that for a given optimal direction the deviation is maximized using all the available impulse capability. This approach provides a fully analytical solution for the maximum deviation CAM in the impulsive case.

The previous method can be extended to the minimization of the collision probability $P$, as first noted by Bombardelli et al. [15] using a different set of generalized orbital elements. Using Chan's series expression for the collision probability, it is possible to prove that minimizing $P$ is equivalent to maximizing:

$$J_P = \delta\mathbf{r}^T\mathbf{Q}^*\delta\mathbf{r} = \delta\mathbf{b}^T\mathbf{Q}^*\delta\mathbf{b}, \quad \text{with } \mathbf{Q}^* = \begin{bmatrix} 1/\sigma_\xi^2 & 0 & -\rho_{\xi\zeta}/\sigma_\xi\sigma_\zeta \\ 0 & 0 & 0 \\ -\rho_{\xi\zeta}/\sigma_\xi\sigma_\zeta & 0 & 1/\sigma_\zeta^2 \end{bmatrix} \quad (17)$$

By substituting the expressions for $\delta \mathbf{b}$ or $\delta \mathbf{r}$, the minimum collision probability problem is reduced to the maximization of a quadratic form of $\delta \mathbf{v}$, which can be solved as an eigenproblem similarly to the maximum deviation case.

## 3.2 Low-thrust CAM analysis

The SA model for constant, tangential, low-thrust CAMs presented in the previous section allows for the efficient computation of the changes in position and velocity at $t_{CA}$, for given values of $a_t$, $t_i$, and $t_e$. This can be leveraged to perform parametric analyses of the possible CAMs, or in combination with an optimization algorithm. Details on the accuracy and numerical performance of the SA formulation are provided in Section 4.

## 3.3 Computation of STMs

Thanks to the inclusion of $\delta \mathbf{r}(t_{CAM})$ in the model for the impulsive manoeuvre and $\delta \mathbf{v}(t_{CA})$ in the relative motion equations, a fully analytical expression for the STM, $\overline{\mathbf{T}} = \mathbf{A} \, \mathbf{G}$, is available for the impulsive case. For other cases, such as those including drag or SRP effects, the STM can be obtained through the numerical propagation of the variational equations using PlanODyn [5], a single-averaged, SA propagator based on Gauss' planetary equations also developed by the COMPASS group. Interestingly, numerical results in Section 4 show that the analytical STM is a good approximation for the propagation of uncertainties even in the presence of drag and SRP. The SA STM for these cases can, however, also be used for the design of CAMs by sails, determining the required effective area-to-mass ratio to achieve a given displacement with respect to the nominal trajectory [1].

## 4 NUMERICAL TEST CASES

The performance of the software tool for different applications is now assessed through several test cases. For brevity, only a small selection of results is presented here; additional details can be found in [1][2][3]. The two different nominal CAs summarized in Table 1 are considered. In all cases the nominal orbital elements of the spacecraft are the same, while two different debris are presented. These values are taken from [1], where an extensive analysis of possible conjunction geometries and resulting optimum CAMs is performed, taking either PROBA-2 or XMM as manoeuvring spacecraft and generating synthetic debris using statistical data from European Space Agency MASTER-2009 model [16].

Table 1. Orbital elements at CA for spacecraft and both debris

|  | $a$ [km] | $e$ [−] | $i$ [deg] | $\Omega$ [deg] | $\omega$ [deg] | $f_0$ [deg] |
|---|---|---|---|---|---|---|
| **Spacecraft (PROBA-2)** | 7093.637 | 0.0014624 | 98.2443 | 303.5949 | 109.4990 | 179.4986 |
| **Debris A** | 7777.097 | 0.0872592 | 70.7378 | 78.1861 | 254.4247 | 7.9912 |
| **Debris B** | 7782.193 | 0.0871621 | 88.6896 | 142.7269 | 248.1679 | 1.2233 |

It is normally difficult to find publicly available information on covariance matrices, as they are provided directly to satellite operators as part of conjunction data messages. For the test cases in this section, the same reference covariance presented in [1][3] is used. This covariance matrix has been estimated from the least-square fitting of SPG4-generated state vectors with the results from a high-accuracy propagator [17], using the two-line elements of catalogue object NORAD ID 33874 (an Iridium 33 debris). The state at orbit determination (OD) is:

$$\mathbf{r}_{OD} = [+6.9687855258 \; 10^{+03} \quad +2.0930747167 \; 10^{+03} \quad -8.0909303360 \; 10^{+00}] \text{ km}$$
$$\mathbf{v}_{OD} = [-1.5353503665 \; 10^{-01} \quad +4.4753975193 \; 10^{-01} \quad +7.3566221447 \; 10^{+00}] \text{ km/s}$$
(18)

and the covariance matrix (units in km and s):

$$C|_{ref} = \begin{bmatrix} +1.155460 \; 10^{-02} & -2.314433 \; 10^{-03} & -1.173196 \; 10^{-03} & +4.525295 \; 10^{-07} & -5.679590 \; 10^{-07} & -1.094546 \; 10^{-05} \\ -2.314433 \; 10^{-03} & +1.914694 \; 10^{-02} & +1.416720 \; 10^{-02} & -1.228650 \; 10^{-05} & -2.553553 \; 10^{-06} & -3.304939 \; 10^{-06} \\ -1.173196 \; 10^{-03} & +1.416720 \; 10^{-02} & +3.087028 \; 10^{-01} & -2.875013 \; 10^{-04} & -8.618777 \; 10^{-05} & -1.249317 \; 10^{-06} \\ +4.525295 \; 10^{-07} & -1.228650 \; 10^{-05} & -2.875013 \; 10^{-04} & +2.885067 \; 10^{-07} & +7.994043 \; 10^{-08} & +1.151141 \; 10^{-09} \\ -5.679590 \; 10^{-07} & -2.553553 \; 10^{-06} & -8.618777 \; 10^{-05} & +7.994043 \; 10^{-08} & +4.599658 \; 10^{-08} & +1.457009 \; 10^{-09} \\ -1.094546 \; 10^{-05} & -3.304939 \; 10^{-06} & -1.249317 \; 10^{-06} & +1.151141 \; 10^{-09} & +1.457009 \; 10^{-09} & +1.202200 \; 10^{-08} \end{bmatrix}$$

The state at OD corresponds to one particular true anomaly. For different true anomalies, the covariance orientation is updated by propagating the reference covariance up to the desired true anomaly and computing the new principal directions, while keeping the volume of the ellipsoid by assigning to these updated principal directions the original eigenvalues at the reference OD [1].

The design of impulsive CAMs is considered first, for the CA between PROBA-2 and Debris A (a detailed analysis for Debris B can be found in [1]). Fig. 2 compares the results obtained for the maximum miss distance and minimum collision probability CAMs, both in terms of miss distance in the b-plane and collision probability, for a $\Delta v$ of 7 m/s, a combined envelope radius of 10 m, and different lead times $\Delta t$ measured in periods of the manoeuvring spacecraft. The differences between both CAMs are due to the orientation of the combined covariance in the b-plane: while the maximum miss distance CAM just tries to increase the deflection as much as possible, the minimum collision probability one also avoids aligning with the principal axis of the covariance ellipse. Because uncertainties tend to grow along the time axis of the b-plane due to phasing effects, and this is also the preferred direction for increasing the total deflection, the minimum collision probability CAM shows strong local minima in miss distance. The differences in collision probability between both solutions are particularly large during the first period, when phasing and geometry change effects can be commensurable. Conversely, phasing effects become dominant as lead time increases and the differences between both solutions in terms of collision probability decrease. This behaviour is easily understood from the b-plane representation in Fig. 3, corresponding to the first local minimum in miss distance for the minimum collision probability CAM, at $\Delta t = 0.6713T$ (for clarity, this $\Delta t$ is also marked in Fig. 2 with grey vertical lines). The minimum $P$ CAMs achieves an appreciably lower collision probability than the maximum separation one despite having a significantly smaller miss distance, because the latter lies along the principal axis of the covariance ellipse while the former is placed perpendicular to it.

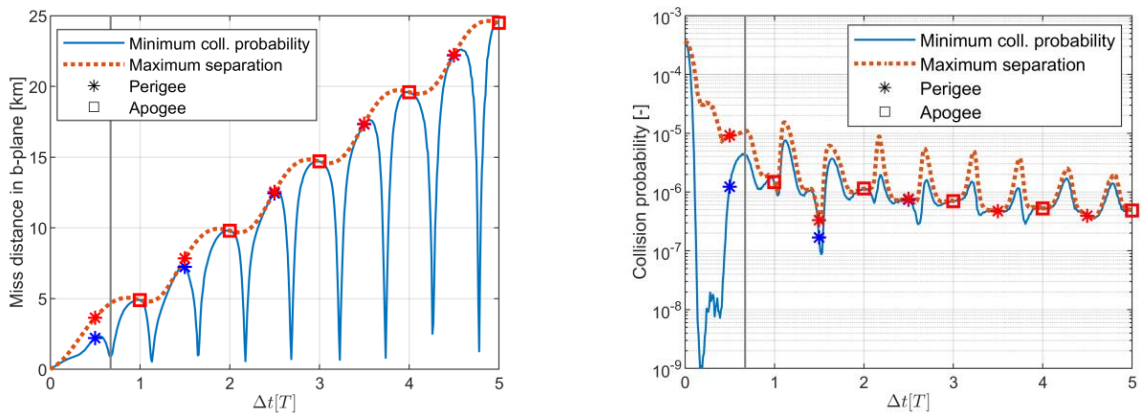

Fig. 2 Miss distance in the b-plane (left) and collision probability (right) for the maximum deviation and minimum collision probability CAMs.

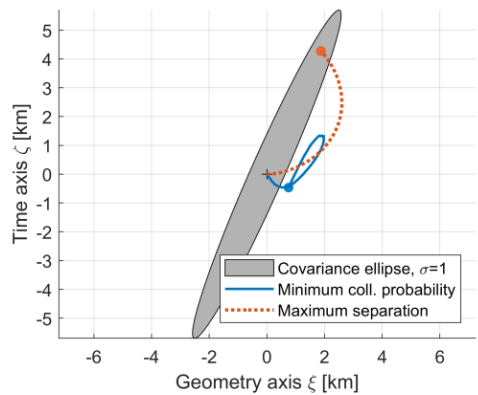

Fig. 3 B-plane representation of the maximum miss distance and minimum collision probability CAMs. The dots represent the optimal CAMs for $\Delta t = 0.6713T$, and the solid lines the traces of the solutions for $\Delta t < 0.6713T$.

Several results for low-thrust CAMs are now considered, for the CA between the spacecraft and Debris B. The deflection in the b-plane, $\delta b$, as a function of $a_t$, the duration of the manoeuvre, $\Delta t_{CAM} = t_e - t_i$, and the time between the end of the CAM and the CA, $\Delta t_f = t_{CA} - t_e$, is characterized in Fig. 4. The synergies between $\Delta t_{CAM}$ and $\Delta t_f$ are appreciated, being possible to increase $\delta b$ for a given $t_{CAM}$ (that is, a given propellant mass) by increasing the coasting arc before the CA. The figure on the right also shows that $\delta b$ follows a nearly linear evolution with $a_t$ in logarithmic scale. This was expected, as the SA model predicts a change in the orbital elements linear with $a_t$, and the displacement is linear in $\delta\boldsymbol{\alpha}$. An erratic behaviour is observed for accelerations of $10^{-3}$ m/s², due to a loss of validity of the SA approach. This is confirmed by the relative errors for $a$, $e$, and $M$ shown in Fig. 5. Although the semi-major axis maintains a high accuracy in all cases, eccentricity and mean anomaly reach errors of up to 10% for $a_t = 10^{-3}$ m/s². However, the errors decrease rapidly with the acceleration, giving good results for practical values of $a_t$ (as reference, a thrust acceleration of $10^{-4}$ m/s² corresponds to a force of 100 mN for a 1 ton spacecraft). The error also depends strongly on the duration of the CAM, which is related to limitations of the short-period correction.

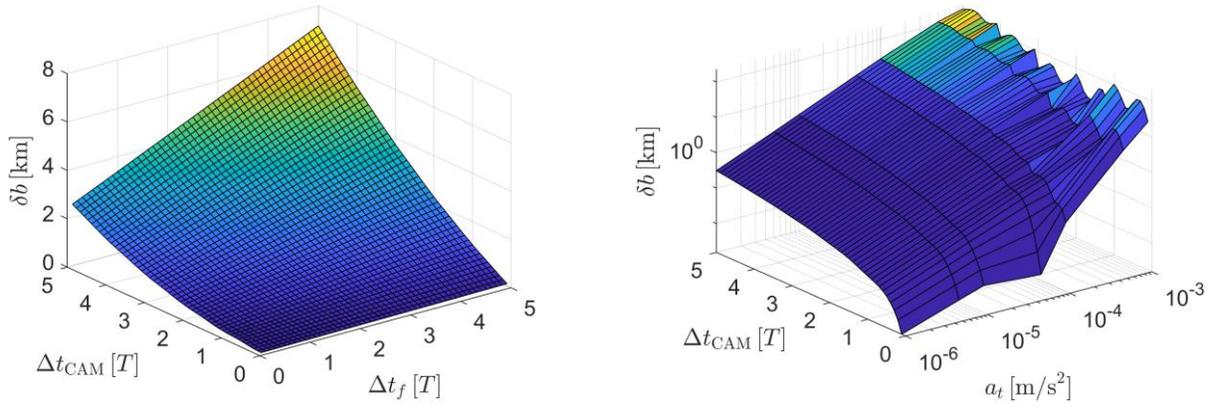

Fig. 4 Deflection in the b-plane as a function of $a_t$, CAM duration $\Delta t_{CAM}$, and time between the end of the CAM and the CA $\Delta t_f$, for a fixed $a_t = 10^{-5}$ m/s²(left) and a fixed $\Delta t_f = T$ (right).

The evolution in the b-plane is characterized in Fig. 6, where each line represents a trace of the CAM in the b-plane as $\Delta t_{CAM}$ increases, for $a_t = 10^{-5}$ m/s² and different values of $\Delta t_f$. As expected, the tangential thrust strategy translates into displacements almost exclusively along the time axis, as its main effect is to change the phasing of the CA. Regarding computational cost, for a numerical campaign comprising 70,000 test cases in a grid with 100 values of $\Delta t_{CAM}$ linearly distributed between 0 and 10 periods of the spacecraft, 100 values of $\Delta t_f$ linearly distributed between 0 and 10 periods, and 7 values of $a_t$ logarithmically distributed between $10^{-3}$ m/s² and $10^{-6}$ m/s², the SA model was 69% faster than the full numerical integration of Gauss' planetary equations. The tests were executed on an Intel Core i7-8700 CPU @3.20 GHz, running Matlab 2019a; to avoid overheads, the code of both the fully numerical and SA methods was compiled using Matlab Coder.

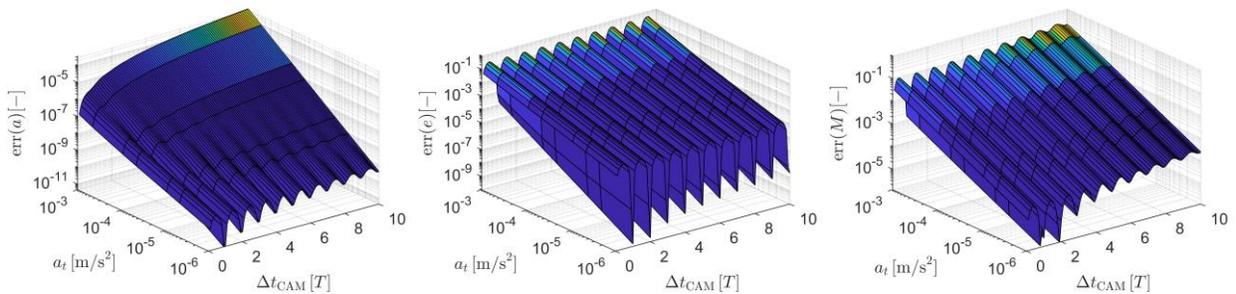

Fig. 5 Relative errors in $a$, $e$, and $M$ of the SA model

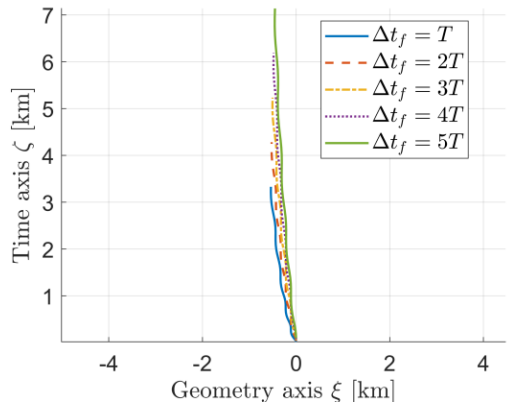

Fig. 6 B-plane representation of the CAM, for $a_t = 10^{-5}$ m/s$^2$

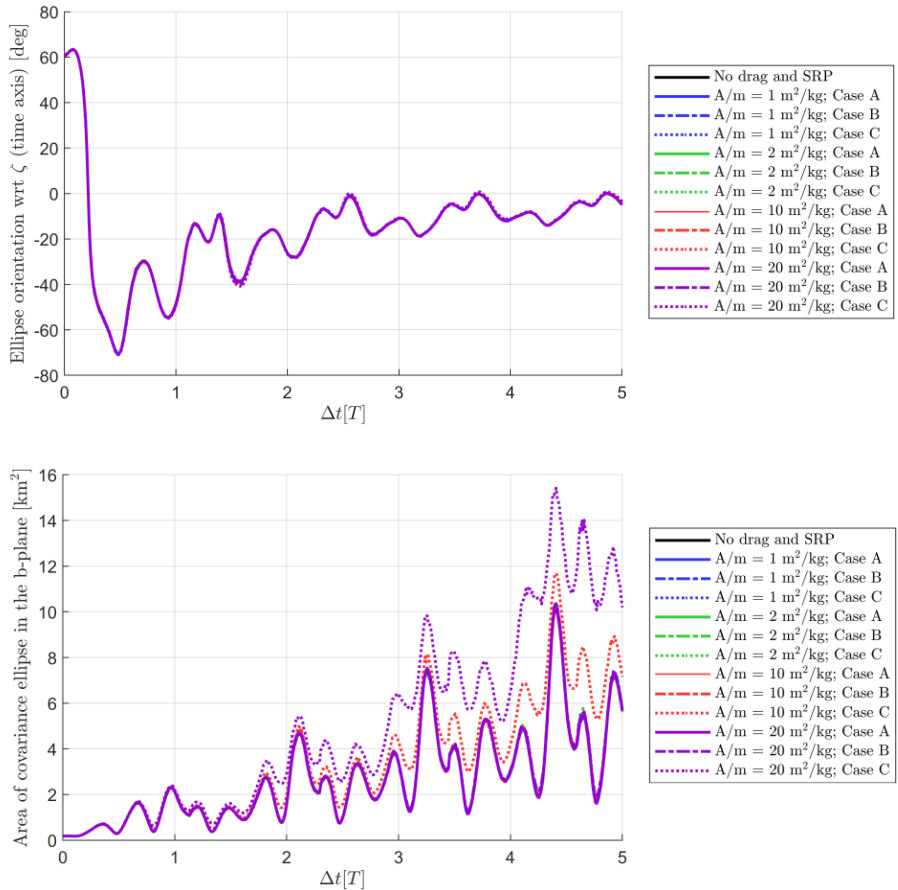

Fig. 7 Orientation with respect to the time axis (top) and area (bottom) of the combined covariance ellipse in the b-plane, for different area-to-mass ratios. Three uncertainty levels are considered for $\sigma_{A/m}$, $\sigma_{c_D}$, and $\sigma_{c_R}$: A) all zero (no uncertainties); B) 1% of their nominal values; C) 10% of their nominal values.

To conclude the numerical test cases, Fig. 7 shows the evolution of size and orientation of the combined covariance ellipse in the b-plane for the spacecraft – Debris B CA, for a range of times between OD and CA, different uncertainties in the reflectivity coefficient $c_R$ and drag coefficient $c_D$, and several area-to-mass ratios. The initial covariances of debris and spacecraft are determined following the procedure detailed at the beginning of this section, the debris has $c_D = 2.1$ and $c_R = 1.8$, and the spacecraft is not affected by drag or SRP. Covariance propagation has been performed numerically through Monte Carlo simulations in PlanODyn, with 10 million elements for each propagation. The foremost conclusion is that the effects of drag and SRP are small except for cases with exceedingly large area-to-mass ratios and uncertainties of the coefficients; therefore, for practical applications they could be neglected and the analytical STM used instead. It is also observed that the covariance tends to align with the time axis of the b-plane as the lead time for the OD grows.

## 5    CONCLUSION

Recent advances in close approach (CA) analysis and collision avoidance manoeuvre (CAM) design using analytical and semi-analytical (SA) methods have been presented. They are part of the software tool MISS (Manoeuvre Intelligence for Space Safety), currently under development by the COMPASS project.

The orbit modifications due to the CAM have been modelled through Gauss' planetary equations. For the impulsive case, a fully analytical linear formulation in the velocity change has been derived, whereas for the low-thrust case, a SA representation has been obtained by assuming tangential thrust. The main limitation of the SA model is that several numerical integrations are still required to get an accurate enough description of the change in phasing at the CA. Nevertheless, numerical tests show a significant reduction in computational time compared with the fully numerical integration of Gauss' planetary equations, and the accuracy of the SA approximation is very good for practical values of the thrust acceleration.

The b-plane representation has been successfully exploited to separate the geometry-change related effects from the phasing-change ones, confirming than the latter dominate over the former as CAM lead time increases. This behaviour is particularly evident for the low-thrust case, where a tangential thrust control law has been imposed. This highlights the importance of planning low-thrust CAMs with sufficient lead time, contrary to impulsive CAMs that can be typically performed in an efficient way up to half an orbital period before the predicted encounter.

MISS also includes analytical procedures for the design of maximum miss distance and minimum collision probability CAMs. The differences between both solutions are due to the orientation of the combined covariance in the b-plane, which tends to align with the time axis as lead time increases (same as the deflected manoeuvre), limiting the reduction in collision probability.

An analytical model for the state transition matrix (STM) in the impulsive case has also been proposed. When other forces such as drag and SRP are present, the STM has been evaluated through numerical integration using PlanODyn, a single-averaged SA propagator. Numerical tests show that drag and SRP have a small effect on the uncertainty evolution, so the analytical STM can be used for uncertainty propagation in most cases.

MISS is under active development, including a deeper exploration of the description of $\delta M$ and the short-periodic corrections of $a$ and $e$, trying to find an approximate, time-explicit solution in order to avoid the numerical parts of the semi-analytical method. Furthermore, PlanODyn capabilities for the native computation of STMs are being extended by implementing the single-averaged Jacobians for more force models.

**DECLARATION OF COMPETING INTEREST**

None.

**FUNDING**

This project has received funding from the European Research Council (ERC) under the European Union's Horizon 2020 research and innovation programme [grant agreement No 679086 – COMPASS].